# The role of crystallographic orientation of martensitic variants on cleavage crack propagation

Arya Chatterjee[a,*], A. Ghosh[a,b], A. Moitra[c], A. K. Bhaduri[c], R. Mitra[a], D. Chakrabarti[a]

[a]Department of Metallurgical and Materials Engineering, Indian Institute of Technology (I.I.T.) Kharagpur, Kharagpur - 721302, West Bengal, India.

[b]Department of Materials Engineering, Indian Institute of Science (IISc), Bengaluru- 560012, Karnataka, India.

[c]Materials Development and Technology Group, Indira Gandhi Center for Atomic Research (IGCAR) Kalpakkam-603102. Tamil Nadu, India.

**Abstract**

Cleavage crack propagation has been investigated in a low-carbon lath-martensitic steel using electron back-scattered diffraction technique. The ability of different martensitic boundaries within prior-austenite grain, such as sub-block, block and packet boundaries to resist cleavage crack propagation has been estimated in terms of Kurdjumov-Sachs crystallographic variants. Crystallographic study of crack path indicated that block boundaries are more effective in cleavage crack deviation as compared to packet boundaries, whilst sub-block boundaries are ineffective in that respect. Moreover, characterizing the boundaries in terms of misorientation angle (angle-axis pair) may be misleading if their effectiveness in retarding cleavage crack propagation is considered.

# 1. Introduction

The modified 9Cr–1Mo grade tempered martensitic steel, used for in-core applications in fast breeder reactors [1] suffers irradiation embrittlement during service, which degrades the impact toughness and increases the ductile–brittle transition temperature (DBTT). In order to improve the impact transition temperature and to avoid catastrophic brittle fracture in unirradiated condition, cleavage crack propagation through the martensitic microstructure of 9Cr-1Mo steel needs to be resisted as far as possible. In order to achieve that the misorientation angle across different boundaries present in martensitic microstructure and the effect of those boundaries on cleavage crack propagation needs to be evaluated, as this aspect is not well-understood.

In case of transgranular fracture of martensitic steels, earlier studies either correlated the size of cleavage facets with the microstructural units such as packet size, block size and lath size [2–4] or estimated the 'effective grain size' by considering only the high-angle boundaries [5–7] as listed in **Table. 1**. In lath martensitic structure, low-angle lath-boundaries are known to be ineffective in resisting the cleavage crack propagation. Packet boundaries and prior-austenite grain boundaries, on the other hand, are generally regarded as high-angle boundaries and effective in resisting cleavage crack propagation.[5,7,8] The role of block boundaries is, however, still unclear from this respect. In this context, the study of crystallographic variants in martensitic structure and their effect on cleavage crack propagation can be useful.[9] So far, the studies dealing with direct observation of cleavage crack path in view of crystallography of martensite are limited.[7,8] Moreover, a recent study showed that the estimation of 'effective grain size' considering grain boundary misorientation angle based on angle-axis pair can be misleading and the angle between the bcc {001} cleavage planes of neighboring crystals need to be taken into account.[10]

The objective of the present study is to evaluate the effectiveness of different sub-structural boundaries such as sub-block, block and packet boundaries within a prior-austenite grain, in resisting the cleavage crack propagation through the martensitic microstructure. In order to achieve the objective, EBSD technique has been used for experimental study of crack propagation through crystallographic variants. The variants were identified from an analytical model, considering K-S orientation relationship during martensitic transformation without knowing the initial orientation of the parent austenite.

## 2. Experimental details

Rolled plates (25 mm thick) of modified 9Cr-1Mo steel containing 0.10 C, 0.41 Mn, 0.21 Si, 0.20 Ni, 8.94 Cr, 0.86 Mo, 0.08 Nb, 0.20 V, 0.05 N and balance Fe (all in wt. %) was received in normalized (1323 K (1050ºC), 25 min) and tempered (1023 K (750ºC), 75 min) condition. In order to generate equiaxed grain structure, a second normalization treatment was given by soaking at 1373 K (1100ºC) for 1 h followed by tempering at 1023 K (750ºC) for 1 h. Instrumented Charpy impact tests were carried out at 77 K (-196ºC) on five standard Charpy V-notch samples (55 mm x 10 mm x 10 mm) prepared along the transverse-longitudinal orientation with respect to the rolled plate. Broken specimens showed complete cleavage fracture, **Figure 1a**, with impact energy absorption of less than 4 J. As the main fracture plane shows the top view of crack path, in order to study the cleavage crack deflection along its propagation the secondary cleavage crack was studied just below the fracture surface on the plane perpendicular to the fracture surface, (**Fig. 1b).** The microstructural study showed tempered martensitic microstructure (**Fig. 1b**) having prior-austenite grain size of 18.6 ± 7.9 μm and martensite packet size of 4.6 ± 1.7 μm estimated from optical and scanning electron micrographs, and lath size of 0.4 ± 0.2 μm measured from transmission electron micrographs (TEM).[11] The details of sample geometry, testing parameters and microstructure have been already reported elsewhere.[12,13] EBSD analysis was carried out on the secondary cleavage crack plane to study the crystal orientation around the crack using HKL Channel 5 system (from Oxford Instruments, UK) fitted in Zeiss® Auriga compact dual beam FIB-FEG microscope operated at a step size of 0.1 μm. EBSD study is preferred over TEM as it can analyse large secondary cleavage crack area with more number of crack-boundary interactions.

3. Arrangement of crystallographic variants and their identification

Kurdjumov-Sachs (K-S) orientation relationship (OR) is generally maintained during austenite to martensitic transformation in low-carbon (< 0.6 wt. %) steel following twenty-four different combinations of K-S variants.[14–16] The arrangement of sub-blocks, having independent K-S variants within a martensitic packet and formation of different types of boundaries within a prior-austenite grain is schematically shown in **Figure 1c**. Among these twenty-four K-S variants, only six variants can be present within a single martensitic packet and those variants are V1-V6, V7-V12, V13-V18 and V19-V24.[14–16] Moreover, three blocks, e.g. V1-V4, V2-V5 and V3-V6 can be formed inside a V1-V6 type packet. A similar hierarchy exists for other types of packets. It is necessary to identify the different K-S variants present within martensitic microstructure to characterize different types of boundaries between them, which will facilitate understanding the role of those boundaries on cleavage crack propagation.

Therefore, an analytical predictive model has been developed for the identification of K-S variants, which neither assumes the orientation of prior-austenite grains nor requires the presence of retained austenite to determine the same. This approach is required as the investigated steel did not show any retained austenite in its service condition. The proposed approach relies on prior-austenite reconstruction based method. From K-S orientation relationship, [9,14,16] the transformation of martensite from austenite can be expressed as:

$$M_{V_i} = T_{V_i} A$$

$$A = T_{V_i}^{-1} M_{V_i} \qquad (1)$$

where, $M_{V_i}$ and $T_{V_i}$ are the orientation and transformation matrix of the i$^{th}$ number of martensitic variant, respectively. 'A' is the orientation matrix of the parent austenite, where $i$=1-24 in K-S OR.

Again from the relation between crystal and sample coordinate system,[17]

$$C_C = g_{\exp} C_S \qquad (2)$$

where, $C_C$ and $C_S$ represent the plane-normal in crystal reference frame and sample reference frame, respectively. $g_{\exp}$ is the orientation of martensitic crystal obtained using EBSD. Now, $g_{\exp} \approx M_{V_i}$ if transformation follows K-S orientation relation. Hence, replacing $M_{V_i}$ with $g_{\exp}$ in equation (1):

$$A = T_{V_i}^{-1}(g_{\exp}) \qquad (3)$$

Let us consider, the orientations of two successive martensitic microstructural units, '$i$' and '$j$', are $g_{\exp-1}$ and $g_{\exp-2}$, respectively. Again, martensitic microstructural unit 'i' and 'j' can be any variant from twenty four possible variant in K-S orientation relationship, for which orientation of austenite will be

$$A_i = T_{V_i}^{-1}(g_{\exp-1}) \qquad \text{for } i\text{=1-24} \qquad (4)$$

$$A_j = T_{V_j}^{-1}(g_{\exp-2}) \qquad \text{for } j\text{=1-24} \qquad (5)$$

Among every possible combinations (i.e. 24 x 24 = 576) of '$i$' and '$j$', there will be one combination for which the deviation ($D_{ij}$) between $A_i$ and $A_j$ is minimum (where $i \neq j$) or close to zero if the 'i' and 'j' unit belong to a single prior austenite. $D_{ij}$ is quantified as misorientation between $A_i$ and $A_j$. Maximum allowable $D_{ij}$ considered in the present study is 3º. Now, for minimum $D_{ij}$, the corresponding 'i' and 'j' have been selected as variants $V_i$ and $V_j$, respectively,. In this way, variants have been identified inside a parent austenite.

## 4. Cleavage angle between martensitic variants

As the cleavage planes of the bcc crystal are {001} type, [5,7,18] there is a possibility of at least three cleavage planes in a crystal for crack propagation. The angle between cleavage planes of two neighboring crystals can be obtained from the following set of mathematical expressions [17]:

$$C_{P1} = g_{exp-1} \, C_{S1}$$

$$C_{S1} = g_{exp-1}^{-1} \, C_{P1} \tag{6}$$

$$C_{S1} = g_{exp-1}^{-1} \begin{bmatrix} 0 \\ 0 \\ 1 \end{bmatrix}_{P1} \tag{7}$$

Similarly, $C_{P2} = g_{exp-2} \, C_{S2}$ (8)

Hence, $C_{S2} = g_{exp-2}^{-1} \begin{bmatrix} 0 \\ 0 \\ 1 \end{bmatrix}_{P2}$ (9)

Now, from equation (7) and (9), $\theta = \cos^{-1}[C_{S1} . C_{S2}]$ (10)

where, $C_P$ and $C_S$ represent the plane normal of the cleavage planes in crystal coordinate system and plane normal of the sample coordinate system, respectively. $g_{exp}$ is the orientation of martensite crystal. Subscripts 1 and 2 refer to 1st and 2nd crystal among the two neighbouring crystals through which the cleavage crack propagates.

There are three possible $C_{P1}$ and $C_{P2}$ (i.e. (100), (010) and (001)) for each martensitic crystal. Hence, there are nine possible combinations of cleavage planes and their corresponding cleavage angles ($\theta$) for the neighbouring crystals. For each $\theta$, there is one projection angle, $\theta_P$, on the surface of the EBSD scan. In case of neighbouring martensitic variants, the combination of cleavage planes for which cleavage crack deflection angle ($\theta_m$), measured from EBSD scanned surface, matches with one of the calculated $\theta_P$, can be regarded as the active cleavage planes. The primary advantage of this model is it determines the active cleavage planes from the orientations obtained directly from EBSD analysis instead of predicting activation of specific cleavage planes as a result of applied loading. A similar approach has been reported by Ghosh et al. [10] in a recent study on ferritic steel.

## 5. Results and discussion

EBSD scans have been performed on several secondary cleavage cracks and around thirty five crack-boundary interactions have been studied to understand the role of different martensitic boundaries on cleavage crack resistance. In this regard, it needs to be mentioned that as the deviation of cleavage crack depends on the relative orientation between two neighboring crystals, hence, the exact orientation of prior-austenite does not play any role when the crack deviation inside a single prior-austenite grain is considered. The general finding is presented below with a few examples. **Figure 2a** shows the path for propagation of a typical ~ 100 μm long secondary crack through the martensitic variants having different orientations.

A part of the crack path shown in **Fig 2a**, where the crack interacts with packet boundary and sub-block boundary has been indicated in **Figure 2b**. First, crack encounters a boundary indicated as B1, where it gets deflected. The adjacent variants have been identified as V10 and V13 of same prior-austenite grain using the model described above. Hence, this boundary can be considered as a packet boundary, as discussed earlier, **Figure 1c**. Next, this crack propagates through a sub-block boundary (B2) between V17 and V14 variants of same prior-austenite grain. The misorientation angles obtained from EBSD for B1 and B2 boundaries are 53.7° and 12.1°, respectively. The calculation shows that {100} and {010} are the active cleavage planes for V10 and V13 variants, respectively, which results in minimum cleavage angle ($\theta_{min}$) of 32.6°. On the other hand, both V17 and V14 variants have {010} as their active cleavage plane and $\theta_{min}$ is 8.2°. However, the measured angles of deviation ($\theta_m$) of the cleavage crack in **Figure 2b** at B1 and B2 boundaries are ~ 40° and ~ 0°, respectively. This discrepancy can be explained by taking into account the projection angles of the cleavage planes on the surface of the observation, i.e. the plane perpendicular to the fracture plane. The calculated projection angles ($\theta_P$) are 42.1° and 0.3° for B1 and B2 boundaries, respectively which marches quite well with the measured angles of deviation ($\theta_m$). Therefore, in this case, the packet boundary effectively causes crack deviation, unlike the sub-block boundary. **Figure 2b** also confirms that the projection angles of the cleavage planes on the plane of observation needs to be considered to determine the exact deviation in cleavage crack path, rather than the angular deviation in crack path directly observed on that plane.

In **Figure 2c**, the crack has been deflected thrice along its propagation at the boundaries between V11 – V9 variants (B3 boundary), V9 – V11 variants (B4-1 boundary) and again at V11 – V9 variants (B4-2 boundary). The misorientation angles at these boundaries are almost same (i.e. ~ 59°), whilst cleavage angle is 41.4° between these variants. At B3 boundary, {010} and {100} are active cleavage planes for V11 and V9 variants, respectively. However, the active cleavage planes are {100} and {001} for V9 and V11 variants, respectively, for both B4-1 and B4-2 boundaries. The cleavage angles ($\theta_m$) at B3, B4-1 and B4-2 are ~43°, ~40° and ~38°, respectively, which are similar to calculated $\theta_P$ of 41°, **Fig. 2c**. As shown in **Figure 1c**, B3, B4-1 and B4-2 boundaries effectively causing cleavage crack deviation are block boundaries, **Figure 2c**.

It is interesting to compare the crack deflection abilities of packet and block boundaries. In **Figure 2d**, the deviation in cleavage crack path has been studied across the B5 boundary, along the interface with V20 and V7 variants. Therefore, according to **Figure 1c**, B5 is a packet boundary, indicating misorientation angle and cleavage angle to be 18.4° and 10.3°, respectively. The calculated projection angle, $\theta_P$ (8°) is similar to the angular deviation, $\theta_m$ (8.5°) observed in **Figure 2d**. The active cleavage planes are found to be {010} and {100} for V20 and V7 variants, respectively. Hence, in spite of being a packet boundary, B5 fails to cause deviation of cleavage crack substantially.

In order to further explain the interaction between the cleavage crack and the boundaries situated inside martensitic structure, a schematic diagram is shown in **Figure 3**. In this case, the actual orientations (Euler angles) of three arbitrary crystallographic variants (V3, V9 and V10) have been chosen with respect to the sample orientation. It is assumed that cleavage crack propagates along the dotted line through three different variants, as indicated by an asterisk in **Figure 1c**. In spite of being a packet boundary, the interface between V3 and V9 variants is expected to cause low-angle (6.07°) deviation of cleavage crack. In contrast, the block boundary between V9 and V10 variants can enforce effective deflection of the crack path by 48.19°.

Crack deviation over 15° threshold angle is generally considered to be high-angle deviation [19]. In the present study, out of more than one hundred interactions between cleavage crack and martensitic boundaries, in 100% cases block boundary has been found to deflect the crack propagation over 15° threshold, considering actual cleavage crack deviation angle. On the other hand, in ~ 75% cases packet boundary lead to crack deviation by 15°. In order to understand this phenomenon, both misorientation (angle-axis pair) and cleavage angles have been theoretically estimated for all possible combinations ($^{24}C_2 = 276$ combinations) of boundaries between K-S variants, **Figure 4**. The possible misorientation angles calculated theoretically as listed in **Figure 4a** show that sub-block boundaries, block boundaries and packet boundaries exceed 15° misorientation in 0%, 100% and 83.33% cases, respectively. The weighted average misorientation angles for sub-block, block and packet boundaries are 10.5°, 57.4° and 40.5°, respectively. The theoretically calculated minimum cleavage angles listed in **Figure 4b** represent that in 0%, 100% and 72.2% cases, high angular deviation (>15°) in cleavage crack path can be expected across the sub-block, block and packet boundaries, respectively. In an earlier work, Guo et al. [9] theoretically showed that the {001} poles of the Bain variants are large angle apart. It is to be noted that the K-S variants which form block boundaries across them are belonging to different Bain variants. Hence, the higher cleavage angle (>15°) across all the block boundaries, presented in theoretical part of the current study is in agreement with the earlier theoretical study of Guo et al. [9]. Further, the present study provides direct experimental evidence of the effectiveness of each type of boundary between the variants. The present study shows that the weighted average angular deviations in cleavage crack paths across sub-block, block and packet boundaries are 7°, 41° and 26°, respectively. The distributions of theoretically calculated misorientation and minimum cleavage angles between all possible combinations of K-S variants, are presented in **Figure 4c** and **d**, respectively. **Figure 4(b, d)** also show that the maximum possible deviation in cleavage crack path within a prior-austenite grain is ~ 48°, which is in-tune with the experimental results. Therefore, the theoretical calculations justify the experimental finding from EBSD analysis, that packet boundaries offer effective barrier in ~75% cases, whilst, block boundaries are always effective from this respect.

## 6. Concluding remarks

The present study has investigated the effect of sub-block, block and packet boundaries on cleavage crack propagation within a lath martensitic structure. It is an established fact that numerous lath boundaries present within sub-block are low-angle boundaries and prior-austenite grain boundaries are high-angle boundaries.[5,7,8] The present investigation indicated that

- block boundaries are more effective in cleavage crack retardation as compared to packet boundaries as all block boundaries are high-angle boundaries considering cleavage angle across the martensitic variants, whilst in ~75% cases packet boundaries offer effective barrier to crack propagation.

- Sub-block boundaries are always low-angle boundaries in terms of angle of misorientation as well as cleavage angle.

- Moreover, the current investigation also recommends that characterization of martensitic boundaries in terms of misorientation angle (angle-axis) and correlating with their ability to resist cleavage crack propagation may be misleading as cleavage occurs on {001} type crystallographic planes in bcc material.

The present model for identifying the crystallographic variants of martensite can also be extended to other OR such as Nishiyama–Wasserman (N-W) and Greninger–Troiano (G–T) etc, where retained austenite is not usually found in the microstructure.

**Acknowledgements**

Authors acknowledge Department of Science and Technology, New Delhi for research grant, SRIC, IIT Kharagpur for equipment grant through SGIRG scheme and IGCAR, Kalpakkam for provision of research material.

**Figure captions:**

**Figure 1**: (a) Fractograph depicting cleavage facets in samples broken at -196 °C, and (b) propagation of secondary cleavage crack across different boundaries (indicated by dotted lines) in martensitic structure; as well as (c) schematic representation of K-S variants surrounded by prior-austenite grain boundary (PAGB). Abbreviation, SBB: sub-block boundary; BB: block boundary; PB: packet boundary.

**Figure 2**: Typical EBSD inverse pole figure maps showing secondary cleavage crack propagation in martensitic structure across different boundaries between martensitic variants.

**Figure 3**: Schematic representation of cleavage crack propagation across three different martensitic variants.

**Figure 4**: List of all possible (a) misorientation angles and (b) cleavage angles between K-S variants, as well as their distribution in (c) and (d), respectively.

**Table caption:**

**Table 1**: Different effective grain considered for cleavage crack resistance and different methods used to identify K-S variants in martensitic steel.

| Ref. | Effective grain | Ref. | Variant identification method |
|---|---|---|---|
| [3,5,8] | Packet | [20,21] | Correlating orientation relationship with existing retained austenite. |
| [18] | 12° misorientation grains | [15,16] | Matching with misorientation angle (angle-axis pair) values |
| [9] | Bain variants | [14] | Correlating with pole figure |
| Present study | Block | Present study | Correlating with orientation of reconstructed prior-austenite |

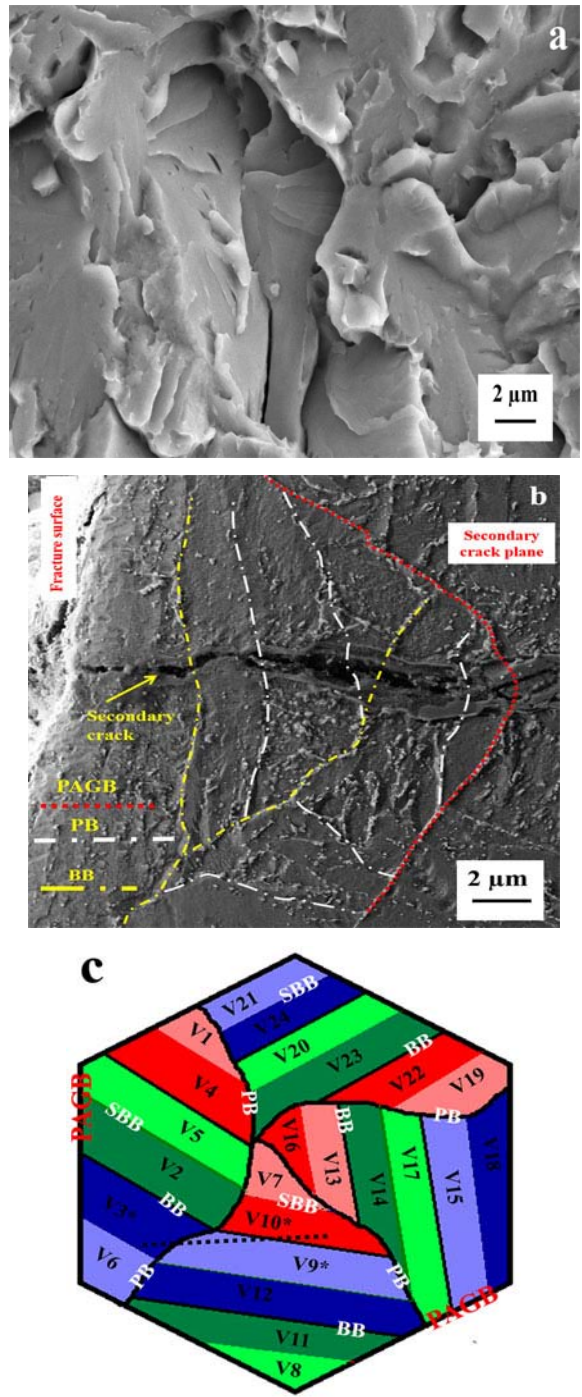

**Figure 1**: (a) Fractograph depicting cleavage facets in samples broken at -196 °C, and (b) propagation of secondary cleavage crack across different boundaries (indicated by dotted lines) in martensitic structure; as well as (c) schematic representation of K-S variants surrounded by prior-austenite grain boundary (PAGB). Abbreviation, SBB: sub-block boundary; BB: block boundary; PB: packet boundary.

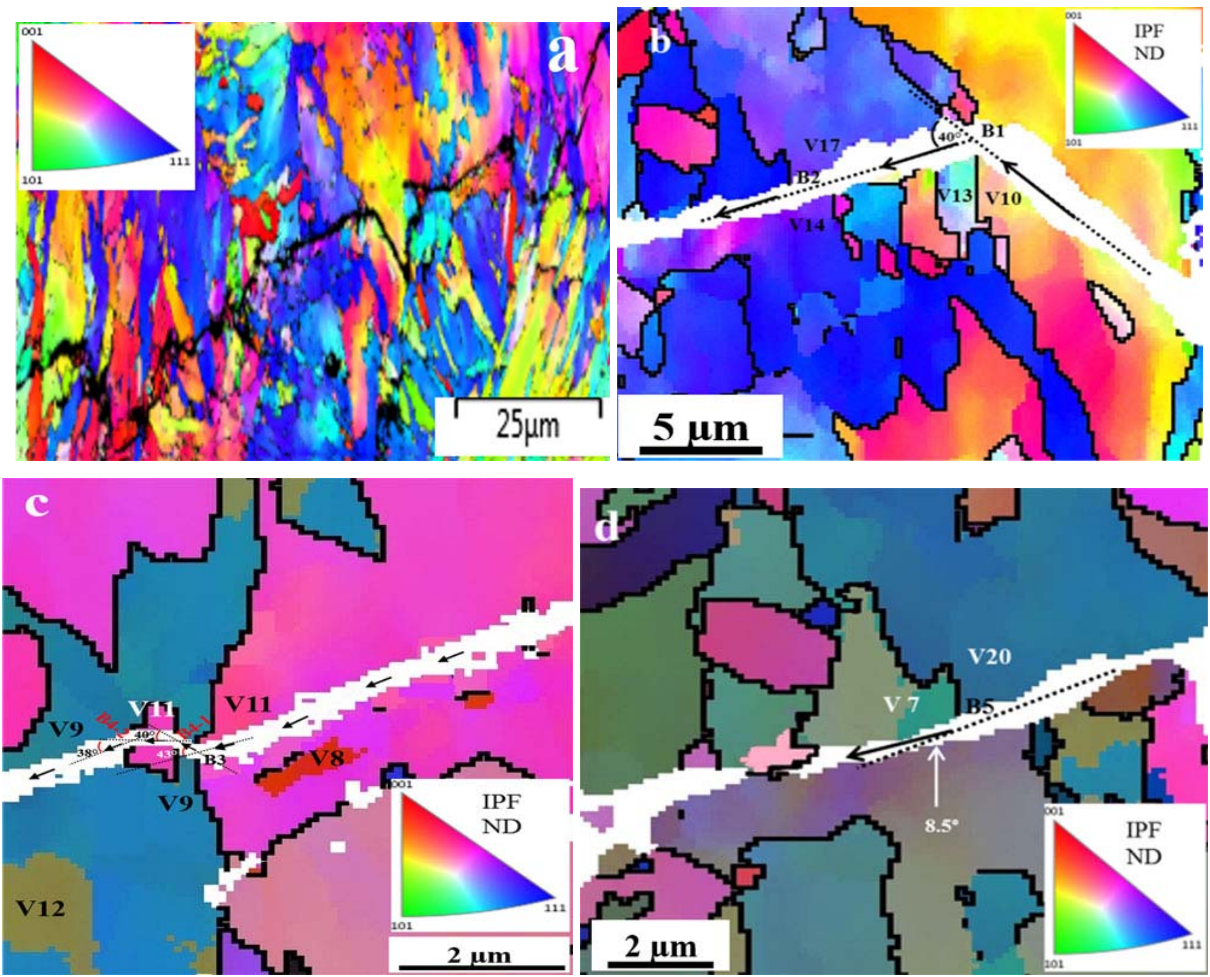

**Figure 2**: Typical EBSD inverse pole figure maps showing secondary cleavage crack propagation in martensitic structure across different boundaries between martensitic variants.

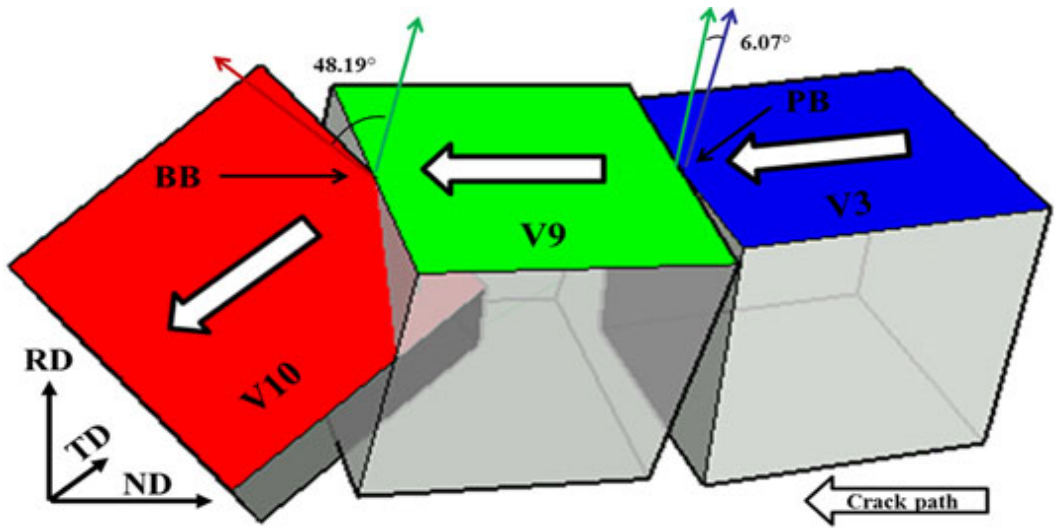

**Figure 3**: Schematic representation of cleavage crack propagation across three different martensitic variants.

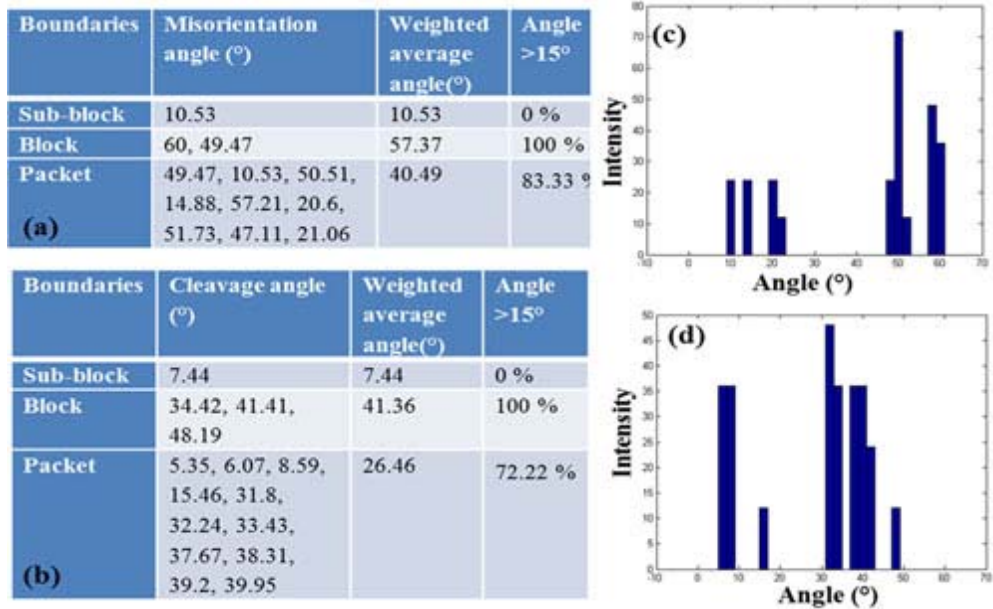

**Figure 4**: List of all possible (a) misorientation angles and (b) cleavage angles between K-S variants, as well as their distribution in (c) and (d), respectively.